\documentclass[onecolumn]{revtex4}
\usepackage{amsmath}
\usepackage{amssymb}

\input{tcilatex}

\begin{document}

\title{Finite-size scaling of partition function zeros and first-order phase
transition for infinitely long Ising cylinder }
\author{${\ }^{1}$Ming-Chang Huang, $^{2}$Tsong-Ming Liaw, $^{1}$Yu-Pin Luo
and $^{2,3}$Simon C. Lin}
\affiliation{$^{1}$Department of Physics, Chung-Yuan Christian University, Chungli
320,Taiwan\\
$^{2}$Computing Centre, Academia Sinica, 11529 Taipei, Taiwan\\
$^{3}$Institute of Physics, Academia Sinica, 11529 Taipei, Taiwan\ }

\begin{abstract}
The critical properties of an infinitely long Ising strip with finite width $%
L$ joined periodically or antiperiodically are investigated by analyzing the
distribution of partition function zeros. For periodic boundary condition,
the the leading finite-size scaling of partition function zeros and its
corrections are given. For antiperiodic boundary condition, the critical
point of $2D$ Ising transition is one of the loci of the zeros, and the
associated non-analyticity is identified as a first-order phase transition.
The exact amount of the latent heat released by the transition is $4/L$.

$\boldsymbol{PACS.}$ 05.50.+q - Lattice theory and statistics.

$\boldsymbol{PACS.}$ 75.10.Hk - Classical spin models.

$\boldsymbol{PACS.}$ 75.70.Cn - Magnetic properties of interfaces.
\end{abstract}

\date{October 19, 2004}
\maketitle


The geometry of the system has been known to play a crucial role in many
aspects of critical behavior: The way of taking the thermodynamic limit may
affect the occurrence of a phase transition, and critical exponents depend
on spatial dimensionality non-trivially. These result in more stringent test
on the universality by including universal critical amplitudes and amplitude
relations obtained from the finite-size effects\cite{privman1,bb}. Boundary
condition is a important factor in determining the finite-size scalings.\ In
particular, aperiodic boundary conditions may introduce interfaces to the
systems and lead to a profound change in the scaling behavior\cite%
{privman2,wumc}.\ In this letter, we report the appearance of a new
first-order transition induced by the interface of a 2D semi-infinite Ising
system. We consider the simplest system, an infinitely long Ising cylinder
defined on a square $L_{x}\times L_{y}$ lattice with infinite $%
L_{x}\rightarrow \infty $ and finite $L_{y}=L$. To have the system with
cylindrical geometry, we impose periodic or antiperiodic condition on the
infinitely extended boundary rows, $\left( m,1\right) $ and $\left(
m,L\right) $ with $-\infty <m<\infty $. The condition on the couplings of
the boundary spin variables $\sigma _{m,L}$ along the $y$-axis is $\sigma
_{m,L+1}=\sigma _{m,1}$ for periodically joined circumference (PJC,
hereafter), and the condition becomes $\sigma _{m,L+1}=-\sigma _{m,1}$ for
antiperiodically joined circumference (AJC, hereafter). The free energy of
the system with AJC contains the additional contribution from the
interfacial tension in comparison with that of the system with PJC\cite%
{lhlw,huang1}. To understand the differences in thermodynamic properties
between the two boundary conditions, we analyze the distribution and the
density of partition function zeros. Our results reveal some intriguing
properties for the system with AJC of finite $L$. \ \ \ \ \ \ \ 

We already have a complete picture about the partition function zeros of the
zero-field square Ising model in the thermodynamic limit\cite{fs,luwu}.
Fisher first showed that the partition function zeros, referred as Fisher
zeros, all lie on the unit circle, $\left\vert z\right\vert =1$, in the
complex plane of $z=\sinh \left( 2\beta \right) $ with the zero on the
positive real axis, $z_{c}=1$, corresponding to the critical point of Ising
transition\cite{fs}. Then, Lu and Wu obtained the density associated with
the zeros on the unit circle analytically\cite{luwu}. Based on these
results, we study the finite-size scalings of Fisher zeros for infinitely
long Ising cylinders. For the system with PJC, portions of the zeros,
including the critical point, are always absent from the unit circle for
finite $L$, and the scaling behavior is given in the exact form. However,
for the system with AJC, regardless the $L$ value, the zeros distribution
always contain the critical point. Then, by analyzing the cumulative
distribution of Fisher zeros around the critical point we show that the
singular property of the critical point can be classified into two types:
One is the first-order transition for a finite circumference, and the other
is the second-order transition for an infinite circumference. The latter
corresponds to the Ising ferromagnetic phase transition, and the former is a
new phase caused by the fluctuation of interface from AJC. Thus, in taking
the thermodynamic limit $L\rightarrow \infty $ from an infinitely long Ising
cylinder of finite $L$, we encounter different situations for two different
boundary conditions: $\left( i\right) $ For the case of PJC, only the
second-order phase transition appears in the limit $L\rightarrow \infty $,
and there is no phase transition in the intermediate stage $0<L<\infty $. $%
\left( ii\right) $ For the case of AJC, we first have the first-order phase
transition which becomes weaker as $L$ increases and eventually reduces to a
second-order phase transition in the limit $L\rightarrow \infty $. This
picture is further verified by measuring the latent heat per site released
in the first-order transition as $4/L$. \ \ \ \ \ \ \ \ 

The partition function of a plane square $L_{x}\times L_{y}$ Ising lattice
with periodic or aperiodic boundary conditions can be expressed as the sum
of four terms\cite{wumc,plechko}, but the sum reduces to one term in the
limit of $L_{x}$ (or $L_{y}$)$\rightarrow \infty $. In this limit, the
system becomes an infinitely long Ising strip with finite width $L$, and the
corresponding free energy of the system per site per $k_{B}T$ is 
\begin{equation}
f_{L}=-\frac{1}{2}\ln \left( 4z\right) -\frac{1}{2L}\sum\limits_{p}%
\int_{0}^{2\pi }\frac{d\phi }{2\pi }\ln \left[ z+\frac{1}{z}-\Phi _{p}(\phi )%
\right] ,
\end{equation}%
for the case of isotropic ferromagnetic couplings\cite{lhlw,huang1}. Here
the possible $p$-values in the sum are half integers ranging from $1/2$ to $%
L-1/2$ for PJC and integers ranging from $0$ to $L-1$ for AJC, and the
function $\Phi _{p}(\phi )$ is given as 
\begin{equation}
\Phi _{p}(\phi )=\cos \phi +\cos \frac{2\pi p}{L}.
\end{equation}%
Then, the loci of the Fisher zeros can be obtained as the union of the
solutions of the condition 
\begin{equation}
z+\frac{1}{z}-\Phi _{p}(\phi )=0
\end{equation}%
with $0\leq \phi \leq 2\pi $ for all allowed $p$-values. The solutions of
Eq. (3) are expressed as 
\begin{equation}
z=\exp \left( \pm i\theta \right) ,
\end{equation}%
for $0\leq \theta \leq \pi $, where $\theta $ are the elements belonging to
the union of the sets of $\theta _{{p}}$ for all allowed $p$-values with 
\begin{equation}
\cos \theta _{{p}}=\frac{\Phi _{p}(\phi )}{2}.
\end{equation}%
For the function $\Phi _{p}(\phi )$ of Eq. (2), the value of $\theta _{{p}}$
for a given ${p}$-mode is in the range between $\theta _{{p,\min }}=\cos
^{-1}\left[ \cos ^{2}\left( \pi p/L\right) \right] $ and $\theta _{{p,\max }%
}=\pi -\cos ^{-1}\left[ \sin ^{2}\left( \pi p/L\right) \right] $. An example
of the curve of $\theta _{{p}}\left( \phi \right) $ obtained from Eq. (5)
for various $p$-modes with $L=6$ is shown in Fig. 1.

For PJC, by including all allowed $p$-modes we obtain the range of $\theta $
as $\theta _{\min }=\theta _{p=1/2,\min }$ and $\theta _{\max }=\pi -\cos
^{-1}\left[ \cos ^{2}\left( \pi /2L\right) \right] $ for even $L$, and $%
\theta _{\min }=\theta _{p=1/2,\min }$ and $\theta _{\max }=\pi $ for odd $L$%
. Thus, as shown in Fig. 2(a) for $L=5$ and 2(c) for $L=6$, the zeros
distribution can be summarized as the followings: $\left( i\right) $ The
zeros are absent from two arcs of the unit circle for even $L$ and from an
arc for odd $L$, and the zeros do not contain the critical point for both
even and odd finite $L$. $\left( ii\right) $ By denoting the nearest zero to
the critical point as $z_{1}\left( L\right) $, we obtain the opening angle
between $z_{1}\left( L\right) $ and the critical point as $\theta _{\min }$.
To obtain the finite-size scaling behavior of $z_{1}\left( L\right) $, we
expand $\sin \theta _{\min }$ and $\cos \theta _{\min }$ as a power series
of $1/L$ to obtain \ 
\begin{equation}
\func{Im}z_{1}\left( L\right) =\sum_{k=0}^{\infty }C_{2k+1}^{I}\left( \frac{%
\pi }{2L}\right) ^{2k+1},
\end{equation}%
with the coefficients $C_{1}^{I}=\sqrt{2}$, $C_{3}^{I}=-5\sqrt{2}/12$, $%
C_{5}^{I}=49\sqrt{2}/480$, and etc.; and 
\begin{equation}
\left\vert \func{Re}z_{1}\left( L\right) -z_{c}\right\vert
=\sum_{k=1}^{\infty }C_{2k}^{R}\left( \frac{\pi }{2L}\right) ^{2k},
\end{equation}%
with the coefficients $C_{2}^{R}=1$, $C_{4}^{R}=-1/3$, $C_{6}^{R}=2/45$, and
etc.. From the finite-size scaling theory of the zeros ditributions, the
leading finite-size scaling behavior of the imaginary part of a Fisher zero
labelled by $j$ is given as 
\begin{equation}
\func{Im}z_{j}\left( L\right) \thicksim L^{-1/\nu },
\end{equation}%
where $\nu $ is the correlation-length exponent\cite{itzykson}; and the
leading scaling behavior of the real part of the lowest zero $\left(
j=1\right) $ can be written as 
\begin{equation}
\left\vert \func{Re}z_{1}\left( L\right) -z_{c}\right\vert \thicksim
L^{-\lambda _{zero}},
\end{equation}%
where $\lambda _{zero}$ is another critical exponent which is closely
related to the shift exponent $\lambda $ characterizing the shift of the
specific-heat peak from the critical point\cite{janke1}. Then, from Eqs. (6)
and (7) we have $\nu =1$ and $\lambda _{zero}=2$. The two values are the
same as the results obtained by Janke and Kenna for a finite rectangular
lattice with the Brascamp-Kunz boundary condition\cite{janke1} but different
from those obtained by other boundary conditions\cite{pirogov}.

For AJC, we obtain the range of $\theta $ as $0\leq \theta \leq \theta
_{\max },$\ \ with $\theta _{\max }=\pi $ for even $L$ and $\theta _{\max
}=\pi -\cos ^{-1}\left[ \cos ^{2}\left( \pi /2L\right) \right] $ for odd $L$%
. Thus, as shown in Fig. 2(b) for $L=5$ and 2(d) for $L=6$, the zeros
distribution is absent from an arc of the unit circle for odd $L$ and fills
the unit circle completely for even $L$. Moreover, the critical point is one
of the loci of the zeros for both even and odd $L$. Then, a question
naturally arises: What is the non-analytic property associated with the zero
at the critical point? This can not be the second-order phase transition
which occurs only in the limit $L\rightarrow \infty $. To determine the
nature of the non-analyticity, we study the functional form of the
cumulative distribution of the zeros\cite{janke2}.

The cumulative distribution of zeros, $G_{L}\left( \theta \right) $, is \
defined as the total number of Fisher zeros in the interval $\left[ 0,\text{ 
}\theta \right] $ of the unit circle. By expressing in terms of the zeros
density $g_{L}\left( \theta \right) $, we have 
\begin{equation}
G_{L}\left( \theta \right) =\int_{0}^{\theta }g_{L}(x)dx.
\end{equation}%
Here we always take the normalization as $G_{L}\left( \pi \right) =1/2$ for
any width $L$. It has been shown\cite{abe,suzuki}that the zeros density $%
g_{\infty }\left( \theta \right) $ near the critical point, $\theta _{c}=0$,
behaves as $g_{\infty }\left( \theta \right) \approx a_{2}\theta ^{1-\alpha
} $ in a second-order phase transition for which the specific heat diverges
as $c\sim \left\vert T-T_{c}\right\vert ^{-\alpha }$. Thence, the
corresponding cumulative distribution of zeros is $G_{\infty }\left( \theta
\right) \approx b_{2}\theta ^{2-\alpha }$. In $2D$ Ising transition for
which we have $\alpha =0$, the density $g_{\infty }\left( \theta \right) $
was shown to vanish linearly as $g_{\infty }\left( \theta \right) \approx
\left\vert \theta \right\vert /2\pi $ near the critical point\cite{luwu},
and the corresponding $G_{\infty }\left( \theta \right) $ obtained by the
integration of $g_{\infty }\left( \theta \right) $ given by Ref. \cite{luwu}
are shown as a solid line in Fig. 3. For the first-order transition, the
asymptotic behavior of $g_{\infty }\left( \theta \right) $ to the critical
point was shown to take the form, $g_{\infty }(\theta )\approx g_{\infty
}(0)+a_{1}\theta ^{n}+...$, with $g_{\infty }(0)$ proportional to the latent
heat\cite{yang2}. Correspondingly, the behavior of $G_{\infty }\left( \theta
\right) $ is 
\begin{equation}
G_{\infty }\left( \theta \right) \approx g_{\infty }(0)\theta +b_{1}\theta
^{n+1}+....
\end{equation}%
\ \ 

For the system of an infinitely long Ising cylinder, the value of $%
G_{L}\left( \theta \right) $ is proportional to the sum of the lengths of
the curves $\theta _{p}\left( \phi \right) $, as shown in Fig. 1, over all
possible $p$-values up to a given $\theta $ value with $0\leq \phi \leq 2\pi 
$. Because of the symmetry, $\theta _{p}\left( \phi \right) =\theta
_{p}\left( -\phi \right) $, we may restrict the $\phi $-values to the
interval $\left[ 0,\pi \right] $. Then, by using the normalization condition 
$G_{L}\left( \pi \right) =1/2$ we write 
\begin{equation}
G_{L}\left( \theta \right) =\frac{1}{2l_{tot}}\sum_{p}l_{p}\left( \theta
\right) ,
\end{equation}%
where $l_{p}\left( \theta \right) $ is the length of the curve $\theta
_{p}\left( \phi \right) $ up to a given $\theta $ value, 
\begin{equation}
l_{p}\left( \theta \right) =\int_{0}^{\phi \left( \theta \right) }\sqrt{%
1+\left( \frac{d\theta _{p}\left( \phi \right) }{d\phi }\right) ^{2}}d\phi ,
\end{equation}%
and $l_{tot}$ is the sum of the lengths of all the curves, 
\begin{equation}
l_{tot}=\sum_{p}l_{p}\left( \theta _{p,\max }\right) .
\end{equation}%
Our numerical result gives $l_{tot}=1.085\left( 2\right) \pi L$. Note that
the value of $l_{tot}$ is linearly proportional to the width $L$ and
independent of boundary conditions. The numerical results of $G_{L}\left(
\theta \right) $ for both PJC and AJC with $L=20$ are shown as circles in
Fig. 3. There are two remarks on Fig. 3 worthy to be mentioned: $\left(
i\right) $ The differences in $G_{L}\left( \theta \right) $ between AJC and
PJC for $L=6$ and $20$ are shown in the up-left corner of Fig. 3, and it
indicates the difference vanishes very quickly as $L$ increases. $\left(
ii\right) $ By comparing with $G_{\infty }\left( \theta \right) $ shown by
the solid line, we know $G_{L}\left( \theta \right) $ converges to $%
G_{\infty }\left( \theta \right) $ very quickly as $L$ increases.

For the system with AJC, the critical point is one of the zeros for both
even and odd $L$, and the cumulative distribution of zeros in the interval $%
0\leq \theta \leq \theta _{{p=1,\min }}$ is contributed solely by the zero
mode $p=0$. In this interval of $\theta $, we have $G_{L}\left( \theta
\right) =l_{p=0}\left( \theta \right) /2l_{tot}$. By completing the
integration of Eq. (13) for $p=0$, we have \ \ \ 
\begin{equation}
G_{L}\left( \theta \right) \approx \left( \frac{\sqrt{3}}{2l_{tot}}\right)
\left( \theta +\frac{1}{36}\theta ^{3}+...\right) .
\end{equation}%
This result yield $g_{L}\left( 0\right) =0.254\left( 0\right) /L$. Thus, by
comparing Eq. (15) with Eq. (11) we may conclude the following properties
for the system with AJC: $\left( i\right) $ The system exhibits a
first-order transition for finite $L$. $\left( ii\right) $ The transition
temperature always locates at the critical point $\beta _{c}$ of $2D$ Ising
transition independent of the value of $L$. $\left( iii\right) $ The latent
heat released by the transition is inversely proportional to the width $L$.
\ \ \ 

To obtain the exact amount of the latent heat released in the transition, we
can measure the discontinuity of the internal energy at the transition
point. The dimensionless internal energy density $\epsilon _{L}$ is defined
as $\epsilon _{L}=\partial f_{L}/\partial \beta $. By rewriting the free
energy density of Eq. (1) as 
\begin{equation}
f_{L}=\ -\frac{1}{4}\ln \left( 4z\right) -\frac{1}{2L}\sum\limits_{p}\left[
-\ln F\left( p,\beta \right) \ +\ I(p,\beta )\right] ,
\end{equation}%
with 
\begin{eqnarray}
I(p,\beta ) &=&\int_{0}^{2\pi }\frac{d\phi }{2\pi }\ln \left[ 1-F(p,\beta
)\cos \phi \right] , \\
F(p,\beta ) &=&z\left( z^{2}+1-\cos \frac{2\pi p}{L}\right) ^{-1},
\end{eqnarray}%
we may obtain $\epsilon _{L}$ by first completing the integration of Eq.
(17) and then performing the derivatives with respect to $\beta $. However,
to avoid any ill-defined result we have to be very cautious of continuity
for the integrand. For conncreteness, we proceed first with the derivative
and then the integration. By using the technique of contour integration, we
have 
\begin{equation}
\frac{\partial I}{\partial \beta }=\left( 1-\frac{1}{\sqrt{1-F^{2}({p},\beta
)}}\right) \,\frac{\partial }{\partial \beta }\ln F({p},\beta ),\quad \text{%
for}\ 1-F^{2}({p},\beta )>0.
\end{equation}%
Note that the above equality holds only when the condition $1-F^{2}({p}%
,\beta )>0$ is met, and this condition ensures the continuity of the
integrand. Then, we have 
\begin{equation}
\epsilon _{L}\ =-\frac{\sqrt{1+z^{2}}}{8z}\ +\ \frac{1}{2L}\sum\limits_{{p}%
}D_{\epsilon }({p},\beta ),
\end{equation}%
with 
\begin{equation}
D_{\epsilon }({p},\beta )=\frac{1}{\sqrt{1-F^{2}({p},\beta )}}\,\frac{%
\partial }{\partial \beta }\ln F({p},\beta ).
\end{equation}

For PJC, the functional values of $F({p},\beta )$ satisfy the inequality $%
0<F(p,\beta )<1$, and the internal energy of Eq. (20) is a continuous
function of $\beta $. For AJC, the functional values of $F({p},\beta )$
also\ satisfy the condition $1-F^{2}({p},\beta )>0$ with one exceptional
point locating at $p=0$ and $\beta =\beta _{c}$ for which we have $1-F^{2}({0%
},\beta _{c})=0$\cite{huang2}. Thus, for AJC we may expect to have the
mismatch of the left and right derivatives of the free energy at $\beta _{c}$%
. The discontinuity $\Delta _{\epsilon }$ defined as the differences between 
$\epsilon _{L}\left( \beta _{c}-\varepsilon \right) $ and $\epsilon
_{L}\left( \beta _{c}+\varepsilon \right) $ exactly is the latent heat per
site released by the transition. Since the zero mode of Eq. (21) is solely
responsible for the discontinuity, we have 
\begin{equation}
\Delta _{\epsilon }=\frac{1}{2L}\lim_{\varepsilon \rightarrow 0}\left[
D_{\epsilon }(0,\beta _{c}-\varepsilon )-D_{\epsilon }(0,\beta
_{c}+\varepsilon )\right] .
\end{equation}%
This yields $\Delta _{\epsilon }=4/L$. Thus, the latent heat scales as $1/L$%
, and this agrees with the result obtained from the analysis of the zeros.

In summary, we analyze the distribution of Fisher zeros in the complex $%
\sinh \left( 2\eta \right) $ plane for an infinitely long Ising strip with
finite width $L$ joined periodically or antiperiodically. For periodically
joined circumferece, the the leading finite-size scaling behavior for the
distribution of Fisher zeros give the exact values of the correlation-length
and shift exponents, $\nu $ and $\lambda _{zero}$, as $\nu =1$ and $\lambda
_{zero}=2$. For antiperiodically joined circumference, the system contains
the interface, and the fluctuation of the interface renders the critical
point of Ising transition to be non-analytic in thermodynamic quantities for
any width $L$. The nature of the non-analytic associated with the critical
point is identified as the first-order phase transition for finite $L$.
Because of the existence of the interface, the system can not be identified
as an effective one-dimensional system with short-range Hamiltonian, and the
appearence of the first-order transition does not contradict with the
Mermin-Wigner theorem\cite{mermin}.The latent heat released by the
transition scales exactly as $1/L$ with the the exact amount $4/L$. Thus, as 
$L$ increases, the strength of the first-order transition decreases, and
eventually the system exhibits the second-order phase transition in the
limit of $L\rightarrow \infty $.

\section{Acknowledgment}

The authors wish to express their gratitude to Prof. V. N. Plechko for
critical reading and constructive comments and to Prof. F. Y. Wu and Dr. M.
C. Wu for stimulated discussions. This work was partially supported by the
National Science Council of Republic of China (Taiwan) under the Grant No.
NSC 91-2112-M-033-008.

\FRAME{ftbpFU}{4.2557in}{3.3849in}{0pt}{\Qcb{The curves of $\protect\theta %
_{p}$ versus $\protect\phi /\protect\pi $ for various $p$-modes with $L=6$
and periodic (solid lines) and antiperiodic (dot lines) boundary conditions.
The number on a curve is the $p$-value}}{}{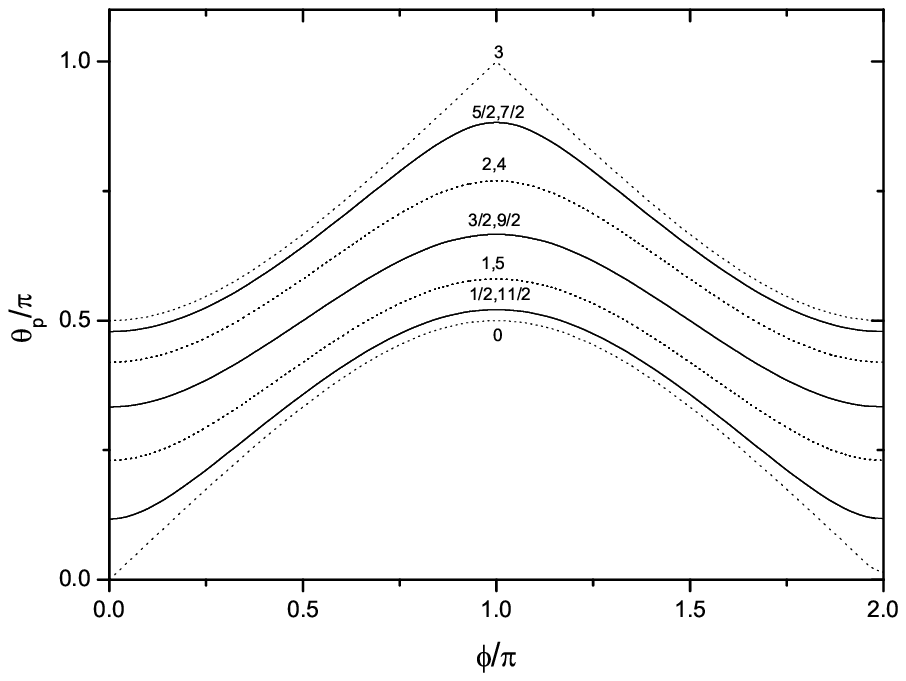}{\special{language
"Scientific Word";type "GRAPHIC";maintain-aspect-ratio TRUE;display
"USEDEF";valid_file "F";width 4.2557in;height 3.3849in;depth
0pt;original-width 4.2056in;original-height 3.3382in;cropleft "0";croptop
"1";cropright "1";cropbottom "0";filename 'fig1.ps';file-properties "XNPEU";}%
}

\FRAME{ftbpFU}{4.2601in}{3.4912in}{0pt}{\Qcb{The distributions of Fisher
zeros in the complex $z$ plane for infinitely long Ising strip with width $%
L=5$ for (a) periodic and (b) antiperiodic boundary conditions and width $L=6
$ for (c) periodic and (d) antiperiodic boundary conditions. The opening
angle is $\protect\theta _{\min }=\cos ^{-1}[\cos ^{2}(\protect\pi /2L)]$}}{%
}{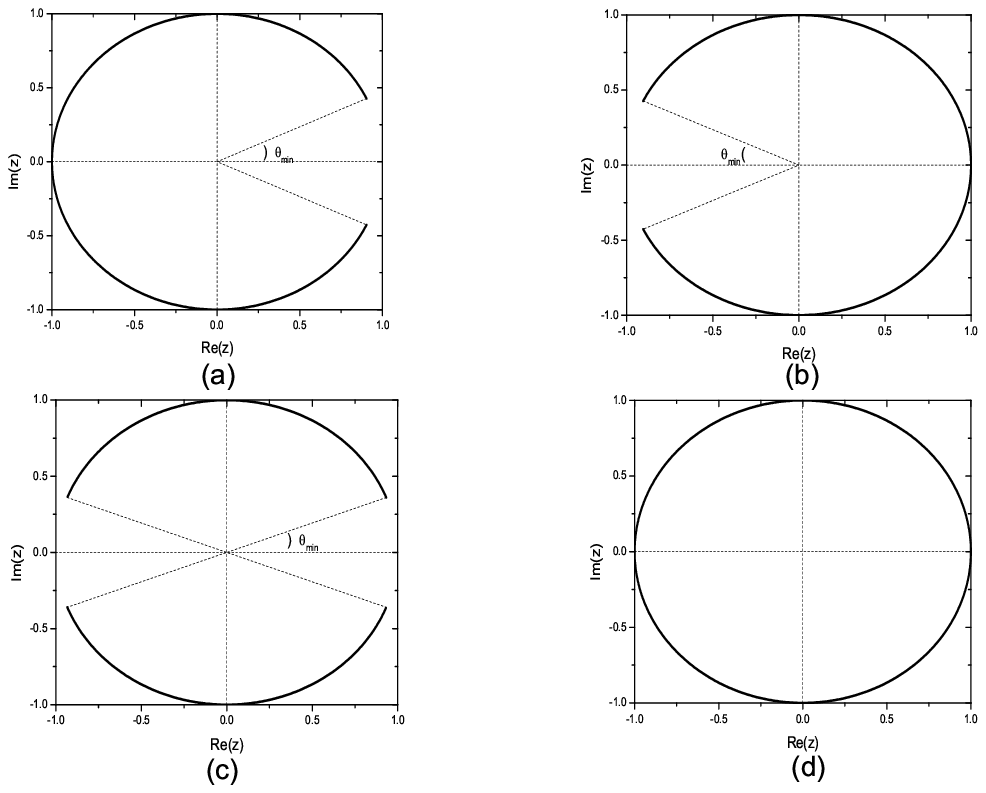}{\special{language "Scientific Word";type
"GRAPHIC";maintain-aspect-ratio TRUE;display "USEDEF";valid_file "F";width
4.2601in;height 3.4912in;depth 0pt;original-width 4.6241in;original-height
3.7844in;cropleft "0";croptop "1";cropright "1";cropbottom "0";filename
'fig2.ps';file-properties "XNPEU";}}

\FRAME{ftbpFU}{4.4399in}{3.2318in}{0pt}{\Qcb{The cumulative distribution of
Fisher zeros $G\left( \protect\theta \right) $ as a function of $\protect%
\theta $. The circles are the results for infinitely long Ising strip with
width $L=20$ for periodic and antiperiodic boundary conditions, and the
solid line corresponds to the results from the integration of the zeros
density given by Ref. $\left[ 8\right] $. The differences in $G_{L}\left( 
\protect\theta \right) $ between the antiperiodic and the periodic boundary
conditions for $L=6$ (dot line) and $20$ (solid line) are shown in the
up-left corner.}}{}{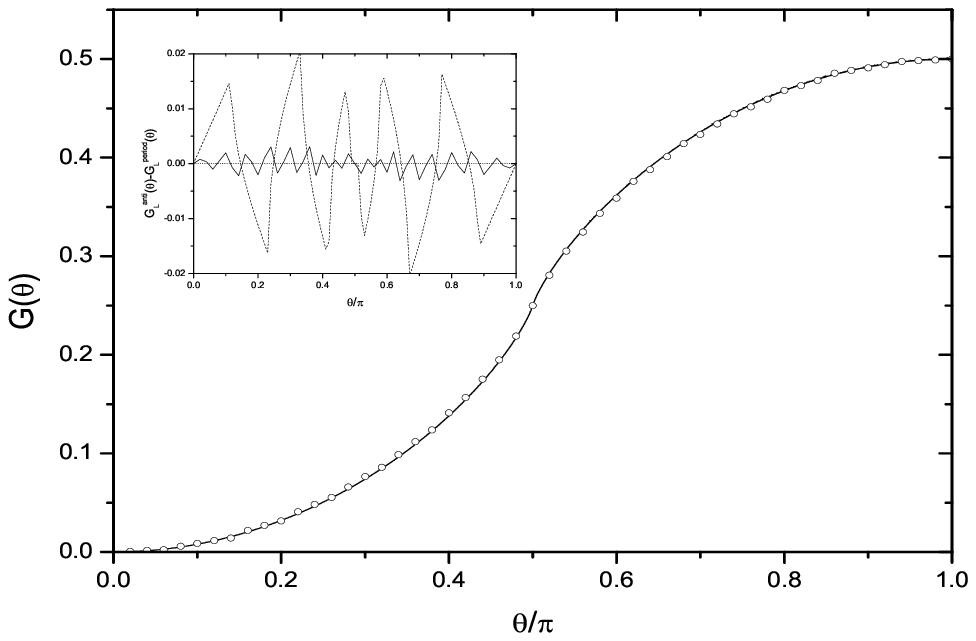}{\special{language "Scientific Word";type
"GRAPHIC";maintain-aspect-ratio TRUE;display "USEDEF";valid_file "F";width
4.4399in;height 3.2318in;depth 0pt;original-width 5.0004in;original-height
3.6322in;cropleft "0";croptop "1";cropright "1";cropbottom "0";filename
'fig3.ps';file-properties "XNPEU";}}

\end{document}